\documentclass{ctr_summer}

\usepackage{ctrfont}

\usepackage[dvips]{graphicx}

\usepackage{psfrag}

\newcommand{\CTRClass}{{\it{\bf ctr\_summer.cls }}}

\title{Magnetic turbulence in compressible fluids}
\shorttitle{MHD turbulence}

\author{Jungyeon Cho
  \footnote{Chungnam National University, Korea} and 
A. Lazarian
  \footnote{University of Wisconsin, USA
    }}

\shortauthor{Cho \& Lazarian}

\begin{document}

\maketitle

We present high resolution numerical simulations
of compressible magnetohydrodynamic (MHD) turbulence.
We concentrate on studies of 
spectra and anisotropy of velocity and density.
We describe a 
technique of separating different magnetohydrodynamic (MHD) modes 
(slow, fast and Alfven) and apply it to our simulations.
For the mildly supersonic case,
velocity and density show anisotropy.
However, for the highly supersonic case, we observe  steepening
of the slow mode velocity spectrum and isotropization of density.
Our present studies show
that the spectrum of density gets substantially
 {\it flatter}
than the Kolmogorov one.

\vskip0.1in
\hrule

\section{Introduction}
Most astrophysical systems, e.g. accretion disks, stellar winds, 
the interstellar medium (ISM) and intercluster medium 
are turbulent 
with an embedded magnetic field that influences almost all
of their properties. This turbulence which spans from km to many
kpc (see discussion in Armstrong, Rickett, \& Spangler 1995)
holds the key to many astrophysical
processes (e.g., transport of mass and angular momentum,
star formation, fragmentation of molecular
clouds, heat and cosmic ray transport, magnetic reconnection).
Statistics of turbulence is also essential for the cosmic microwave
background (CMB) radiation foreground
studies (Lazarian \& Prunet 2002).
In this brief, using high resolution simulations,
we discuss statistics of 3D MHD turbulence and
present new results on density fluctuations.

 Why do we expect astrophysical fluids to be turbulent?
A fluid of viscosity $\nu$ becomes turbulent when the rate of viscous 
dissipation, which is  $\sim \nu/L^2$ at the energy injection scale $L$, 
is much smaller than
the energy transfer rate $\sim V_L/L$, where $V_L$ is the velocity dispersion
at the scale $L$. The ratio of the two rates is the Reynolds number 
$Re=V_LL/\nu$. In general, when $Re$ is larger than $10-100$
the system becomes turbulent. Chaotic structures develop gradually as 
$Re$ increases,
and those with $Re\sim10^3$ are appreciably less chaotic than those
with $Re\sim10^8$. Observed features such as star forming clouds are
very chaotic for $Re>10^8$. 
This makes it difficult to simulate realistic turbulence. 
The currently available
3D simulations containing 512 grid cells along each side
can have $Re$ up to $\sim O(10^3)$
and are limited by their grid sizes. 
Therefore, it is essential to find ``{\it scaling laws}" in order to
extrapolate numerical calculations ($Re \sim O(10^3)$) to
real astrophysical fluids ($Re>10^8$). 
We show below that even with its limited resolution, numerics is a great 
tool for {\it testing} scaling laws.

Kolmogorov theory provides a scaling law for {\it incompressible} 
{\it non}-magnetized hydrodynamic turbulence.
This law provides a statistical relation
between the relative velocity $v_l$ of fluid elements and their separation
$l$, namely, $v_l\sim l^{1/3}$.  An equivalent description is to 
express spectrum $E(k)$
as a function of wave number $k$ ($\sim 1/l$).
The two descriptions are related by $kE(k) \sim v_l^2$. The famous
Kolmogorov spectrum is  $E(k)\sim k^{-5/3}$. The applications of 
Kolmogorov theory range from engineering research to
meteorology (see Monin \& Yaglom 1975) but its astrophysical
applications are poorly justified and the application
of the Kolmogorov theory can lead to erroneous conclusions
(see reviews by Lazarian {\it et al.}
2003; Lazarian \& Yan 2003)

Let us consider {\it incompressible} MHD turbulence first.
There have long been an understanding that the MHD turbulence
is anisotropic
(e.g. Shebalin {\it et al.}~1983). Substantial progress has been achieved
recently by Goldreich \& Sridhar (1995; hereafter GS95), who made an
ingenious prediction regarding relative motions parallel and
perpendicular to magnetic field {\bf B} for incompressible
MHD turbulence. 
An important observation that leads to understanding of the GS95
scaling is that magnetic field cannot prevent mixing motions
of magnetic field lines if the motions
are perpendicular to the magnetic field. Those motions will cause, however,
waves that will propagate along magnetic field lines.
If that is the case, 
the time scale of the wave-like motions along the field, 
i.e. $\sim l_{\|}/V_A$,
($l_{\|}$ is the characteristic size of the perturbation along 
the magnetic field and 
$V_A=B/\sqrt{4 \pi \rho}$ is 
the local Alfven speed) will be equal to the hydrodynamic time-scale, 
$l_{\perp}/v_l$, 
where $l_{\perp}$ is the characteristic size of the perturbation
perpendicular to the magnetic field.
The mixing motions are 
hydrodynamic-like\footnote{
        Recent simulations (Cho {\it et al.}~2003) suggest that
        perpendicular mixing is indeed efficient for
        mean magnetic fields of up to the equipartition value.}.
They obey Kolmogorov scaling,
$v_l\propto l_{\perp}^{1/3}$, because incompressible turbulence is assumed. 
Combining the two relations above
we can get the GS95 anisotropy, $l_{\|}\propto l_{\perp}^{2/3}$ 
(or $k_{\|}\propto k_{\perp}^{2/3}$ in terms of wave-numbers).
If  we interpret $l_{\|}$ as the eddy size in the direction of the 
local 
magnetic field
and $l_{\perp}$ as that in the perpendicular directions,
the relation implies that smaller eddies are more elongated.

GS95 predictions have been confirmed 
numerically (Cho \& Vishniac 2000; Maron \& Goldreich 2001;
Cho, Lazarian \& Vishniac 2002a, hereafter CLV02a; 
see also CLV03); 
they are in good agreement with observed and inferred astrophysical spectra 
(see CLV03). However, the
GS95 model considered incompressible MHD, but the media in
molecular clouds is {\it highly compressible}. Does any part
of GS95 model survives?
Literature on the properties of compressible MHD is very rich (see reviews
by Pouquet 1999; Cho \& Lazarian 2004 and references therein).
Higdon (1984) theoretically studied density fluctuations
in the interstellar MHD turbulence.
Matthaeus \& Brown (1988) studied nearly incompressible MHD at low Mach
number and Zank \& Matthaeus (1993) extended it. In an important paper
Matthaeus {\it et al.}~(1996) numerically
explored anisotropy of compressible MHD turbulence. However, those
papers do not provide universal scalings of the GS95 type.
After the GS95 model, Lithwick \& Goldreich studied scaling
relations for high-$beta$ plasmas 
and Cho \& Lazarian (2002; hereafter CL02)
for low-$beta$ plasmas.

The complexity of the
compressible magnetized turbulence with magnetic field made some
researchers believe that the phenomenon is too complex to expect any
universal scalings for molecular cloud research.
High coupling of compressible
and incompressible motions is often quoted to justify this 
point of view.

Below we shall provide arguments that are suggestive that the fundamentals
of compressible MHD can be understood and successfully applied to
astrophysical fluids.
In many astrophysical fluids the regular (or {\it uniform}) 
magnetic field is comparable
with the fluctuating one. Therefore 
for most part of our discussion, we shall discuss results
obtained for 
$\delta V \sim \delta B/\sqrt{4 \pi \rho} \sim B_0/\sqrt{4 \pi \rho}$,
where $\delta B$ is the r.m.s. strength of the random magnetic field.

Our work done during the summer program at CTR that we report here
was an attempt to get a better insight into the physics of MHD turbulence
using high resolution simulations (i.e. simulations within a $512^3$ box).
In what follows we discuss our numerical approach (sect. 2), our
results obtained for the velocity and magnetic field statistics (sect. 3)
and the density statistics (sect. 4). We concentrate on studies of 
spectra and turbulence anisotropy.

\section{Numerical approach}
We use a third-order accurate hybrid
essentially non-oscillatory (ENO) scheme (see CL02) 
to solve the ideal isothermal
MHD equations in a periodic box:
\begin{eqnarray}
{\partial \rho    }/{\partial t} + \nabla \cdot (\rho {\bf v}) =0,  \\
{\partial {\bf v} }/{\partial t} + {\bf v}\cdot \nabla {\bf v} 
   +  \rho^{-1}  \nabla(a^2\rho)
   - (\nabla \times {\bf B})\times {\bf B}/4\pi \rho ={\bf f},  \\
{\partial {\bf B}}/{\partial t} -
     \nabla \times ({\bf v} \times{\bf B}) =0, 
\end{eqnarray}
with
    $ \nabla \cdot {\bf B}= 0$ and an isothermal equation of state.
Here $\bf{f}$ is a random large-scale driving force, 
$\rho$ is density,
${\bf v}$ is the velocity,
and ${\bf B}$ is magnetic field.
The rms velocity $\delta V$ is maintained to be approximately unity, 
so that ${\bf v}$ can be viewed as the velocity 
measured in units of the r.m.s. velocity
of the system and ${\bf B}/\sqrt{4 \pi \rho}$ 
as the Alfv\'{e}n velocity in the same units.
The time $t$ is in units of the large eddy turnover time ($\sim L/\delta V$) 
and the length in units of $L$, the scale of the energy injection.
The magnetic field consists of the uniform background field and a
fluctuating field: ${\bf B}= {\bf B}_0 + {\bf b}$.

We drive turbulence solenoidally in Fourier space and
use $512^3$ points, $V_A=B_0/\sqrt{4 \pi \rho}=1$, and $\rho_0=1$. 
The average rms velocity in statistically stationary state, $\delta V$,
is around $1$.

For our calculations we assume that
$B_0/\sqrt{4 \pi \rho} \sim \delta B/\sqrt{4 \pi \rho} \sim \delta V$.
In this case, the sound speed is the controlling parameter and
basically two regimes can exist: supersonic and subsonic.
Note that supersonic means low-beta and subsonic means high-beta.
When supersonic, we consider mildly supersonic (or, mildly low-$\beta$)
and highly supersonic (or, very low-$\beta$)\footnote{
        The terms ``mildly'' and ``highly''  are 
        somewhat arbitrary terms.
        We consider these two supersonic cases to cover
        a broad range of parameter space.
        Note that Boldyrev, Nordlund, \& Padoan (2002)
        recently provided a Mach number dependence study of the 
        compressible MHD turbulence statistics where only two regimes 
        are manifest: essentially incompressible and essentially 
        compressible shock-dominated (with smooth transition at some $M_s$ of 
        order unity).
}.

\subsection{Separation of MHD modes}    \label{section_decomposition}
Three types of waves exist (Alfven, slow and fast)
in compressible magnetized plasma. 
The slow, fast, and Alfven bases that denote the direction of displacement
vectors for each mode are given by
\begin{eqnarray}
   \hat{\bf \xi}_s \propto 
     ( -1 + \alpha - \sqrt{D} )
            k_{\|} \hat{\bf k}_{\|} 
     + 
     ( 1+\alpha - \sqrt{D} ) k_{\perp} \hat{\bf k}_{\perp},
  \label{eq_xis_new}
\\
   \hat{\bf \xi}_f \propto 
     ( -1 + \alpha + \sqrt{D} )
           k_{\|}  \hat{\bf k}_{\|} 
     + 
     ( 1+\alpha + \sqrt{D} ) k_{\perp} \hat{\bf k}_{\perp},  
   \label{eq_xif_new}
\\
 \hat{\bf \xi}_A = -\hat{\bf \varphi} 
         = \hat{\bf k}_{\perp} \times \hat{\bf k}_{\|},
\end{eqnarray}
where $D=(1+\alpha)^2-4\alpha \cos\theta$, $\alpha=a^2/V_A^2=\beta(\gamma/2)$,
$\theta$ is the angle between ${\bf k}$ and ${\bf B}_0$, and
$\hat{\bf \varphi}$ is the azimuthal basis in the spherical polar coordinate
system.
These are equivalent to the expression in CL02:
\begin{eqnarray}
   \hat{\bf \xi}_s &\propto &
        k_{\|} \hat{\bf k}_{\|}+
     \frac{ 1-\sqrt{D}-{\beta}/2  }{ 1+\sqrt{D}+{\beta}/2  } 
    \left[ \frac{ k_{\|} }{ k_{\perp} }  \right]^2
     k_{\perp} \hat{\bf k}_{\perp},  \label{eq_xis}     \\
   \hat{\bf \xi}_f &\propto &
     \frac{ 1-\sqrt{D}+{\beta}/2  }{ 1+\sqrt{D}-{\beta}/2  } 
    \left[ \frac{ k_{\perp} }{ k_{\|} } \right]^2
     k_{\|} \hat{\bf k}_{\|}  +
          k_{\perp} \hat{\bf k}_{\perp}.
\end{eqnarray}
(Note that $\gamma=1$ for isothermal case.)

Slow and fast velocity components can be obtained 
by projecting velocity Fourier component 
${\bf v}_{\bf k}$ onto $\hat{\bf \xi}_s$ and $\hat{\bf \xi}_f$, respectively.
See Cho \& Lazarian (2003; hereinafter CL03) for discussion regarding 
how to separate slow and fast magnetic modes.
We obtain energy spectra using this projection method.
When we calculate 
the structure functions (e.g.~for Alfv\'{e}n modes)
in real space, 
we first obtain the Fourier components using the projection and, then, 
we obtain the real space values by performing Fourier transform.

\begin{figure*}
  \includegraphics[width=0.90\textwidth]{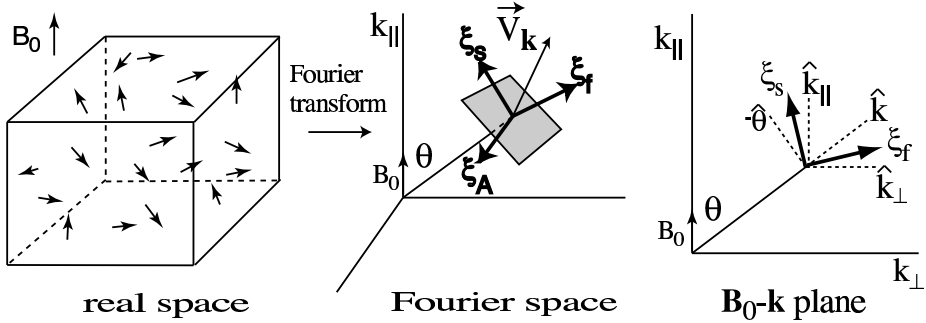}
  \caption{
      Separation method. We separate Alfven, slow, and fast modes in Fourier
      space by projecting the velocity Fourier component ${\bf v_k}$ onto
      bases ${\bf \xi}_A$, ${\bf \xi}_s$, and ${\bf \xi}_f$, respectively.
      Note that ${\bf \xi}_A = -\hat{\bf \varphi}$. 
      Slow basis ${\bf \xi}_s$ and fast basis ${\bf \xi}_f$ lie in the
      plane defined by ${\bf B}_0$ and ${\bf k}$.
      Slow basis ${\bf \xi}_s$ lies between $-\hat{\bf \theta}$ and 
      $\hat{\bf k}_{\|}$.
      Fast basis ${\bf \xi}_f$ lies between $\hat{\bf k}$ and 
      $\hat{\bf k}_{\perp}$. From Cho \& Lazarian (2003).
}
\label{fig_separation}
\end{figure*}

We tested our technique of separation in CL03 for a case when
the separation is possible in real space and got essentially identical
results with our {\it statistical} technique. Therefore we believe that
our separation procedure works reliably. 

\section{Velocity and magnetic field spectra}

We show in Fig.~2 new results from high resolution simulation of highly supersonic magnetically dominated media.
The sonic Mach number, $M_s$,  is $\sim 10$.
Fig. 2 shows that most of the scaling relations we previously found
in the low resolution simulations are still valid
in the high resolution simulation.
Especially anisotropy of Alfven, slow, and fast modes
is almost identical to the one in the previous studies.
However, the power spectra for slow modes do not show
the Kolmogorov slope.
The slope is close to $-2$, which is suggestive of shock formation.
At this moment, it is not clear whether or not the $-2$ slope
is the true slope.
In other words, the observed $-2$ slope might be due to the limited
numerical resolution.
Runs with higher numerical resolution should
give the definite answer.

Formation of shocks is expected as within slow modes for sufficiently
high Mach numbers the turbulence gets superAlfvenic. However, it is
interesting that the turbulence anisotropy is definitely affected by
magnetic field.

Alfven modes follow the GS95 scaling as in the incompressible 
MHD turbulence.
Alfven perturbations cascade to small scales over just one wave
period, while the other non-linear interactions require more time.
Therefore we expect
that the non-linear interactions with other types of waves
should affect Alfvenic cascade only marginally. 
Moreover, as the Alfven waves are incompressible, the properties
of the corresponding cascade do not depend on the sonic Mach number.

\begin{figure*}
\begin{tabbing}
\includegraphics[width=.33\textwidth]{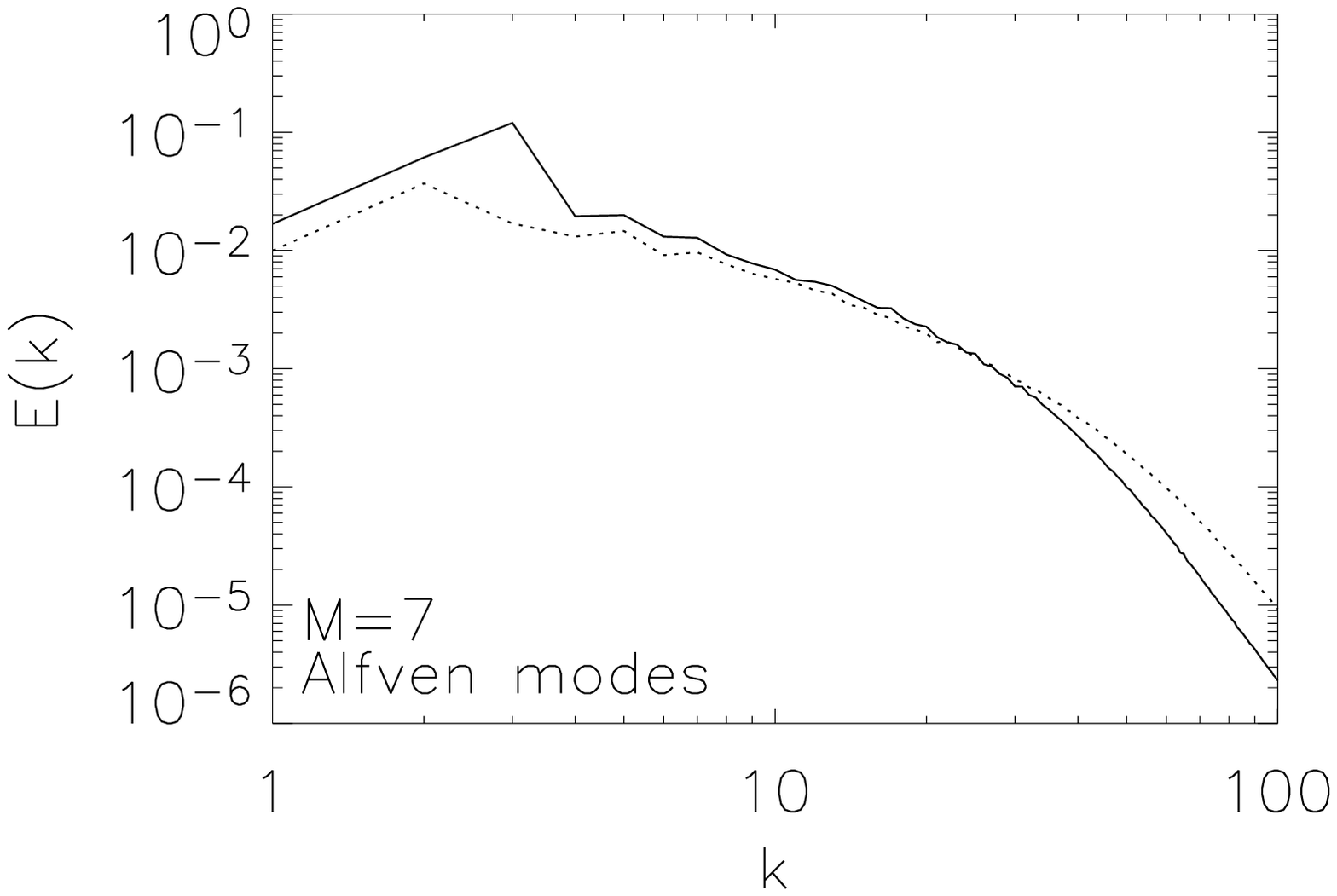}
\=  
\includegraphics[width=.33\textwidth]{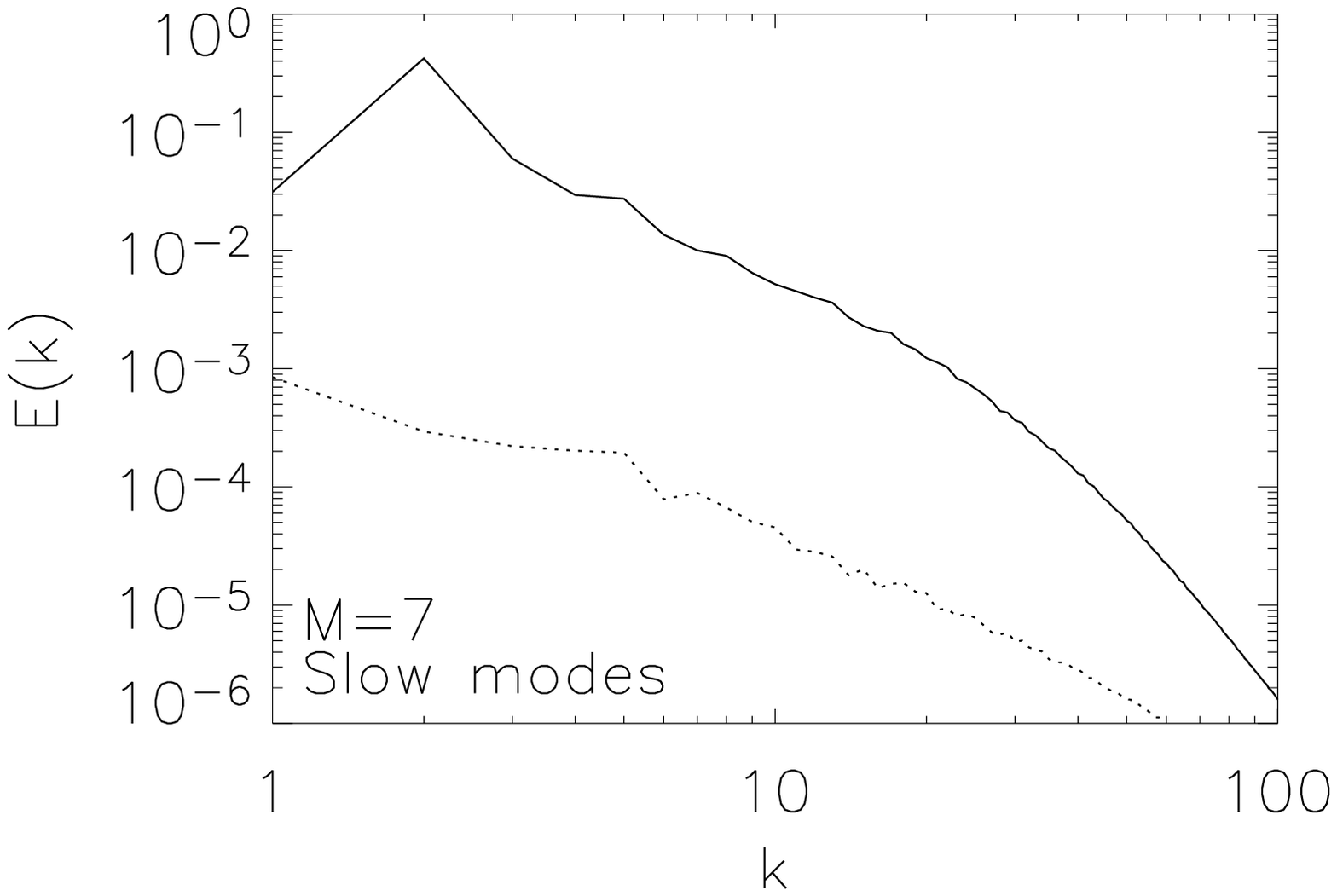}
\=  
\includegraphics[width=.33\textwidth]{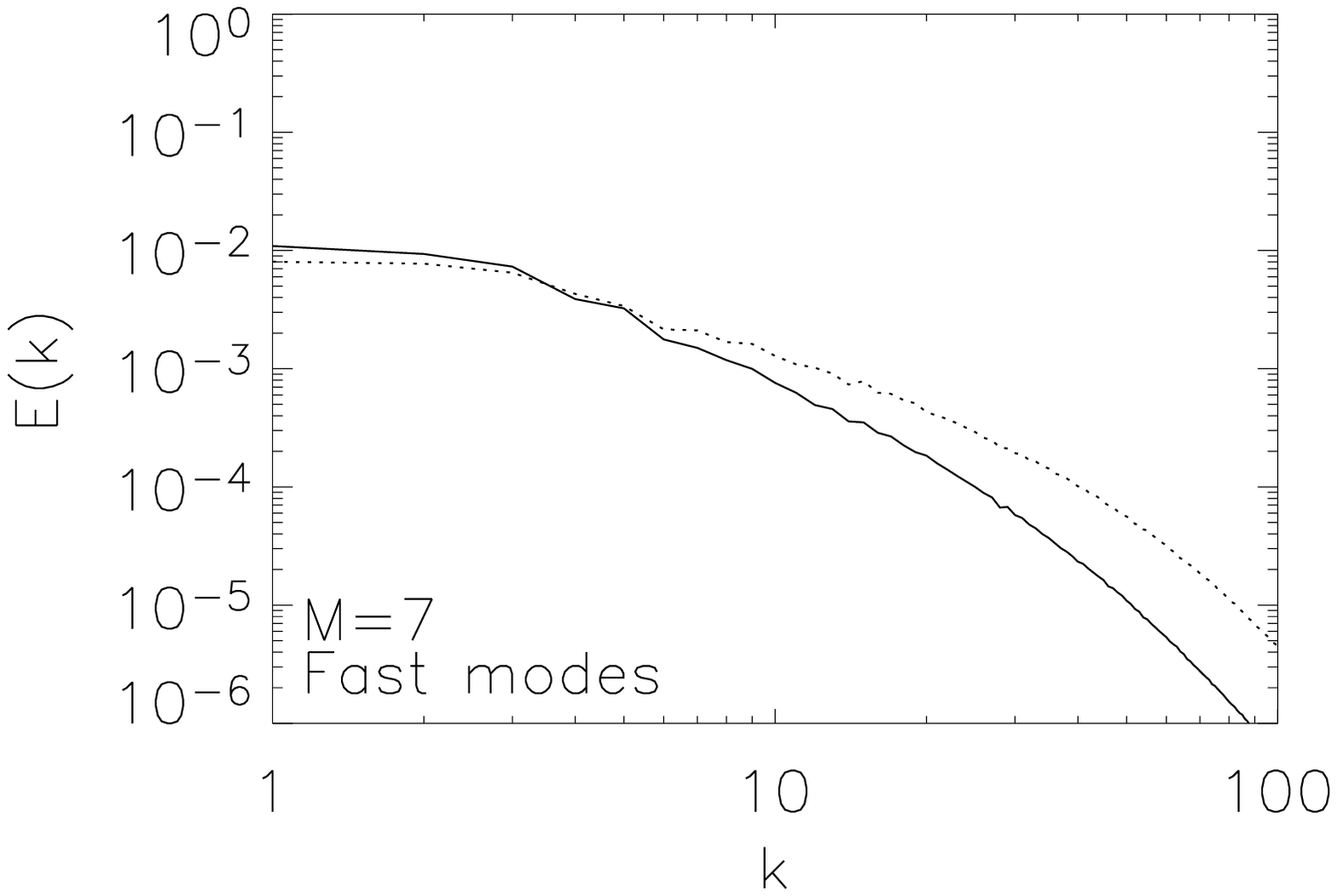}
\\   
 \hspace{23mm} (a) \= \hspace{39mm}  (c)   \hspace{40mm} \=   (e)
\\
\\
\includegraphics[width=.33\textwidth]{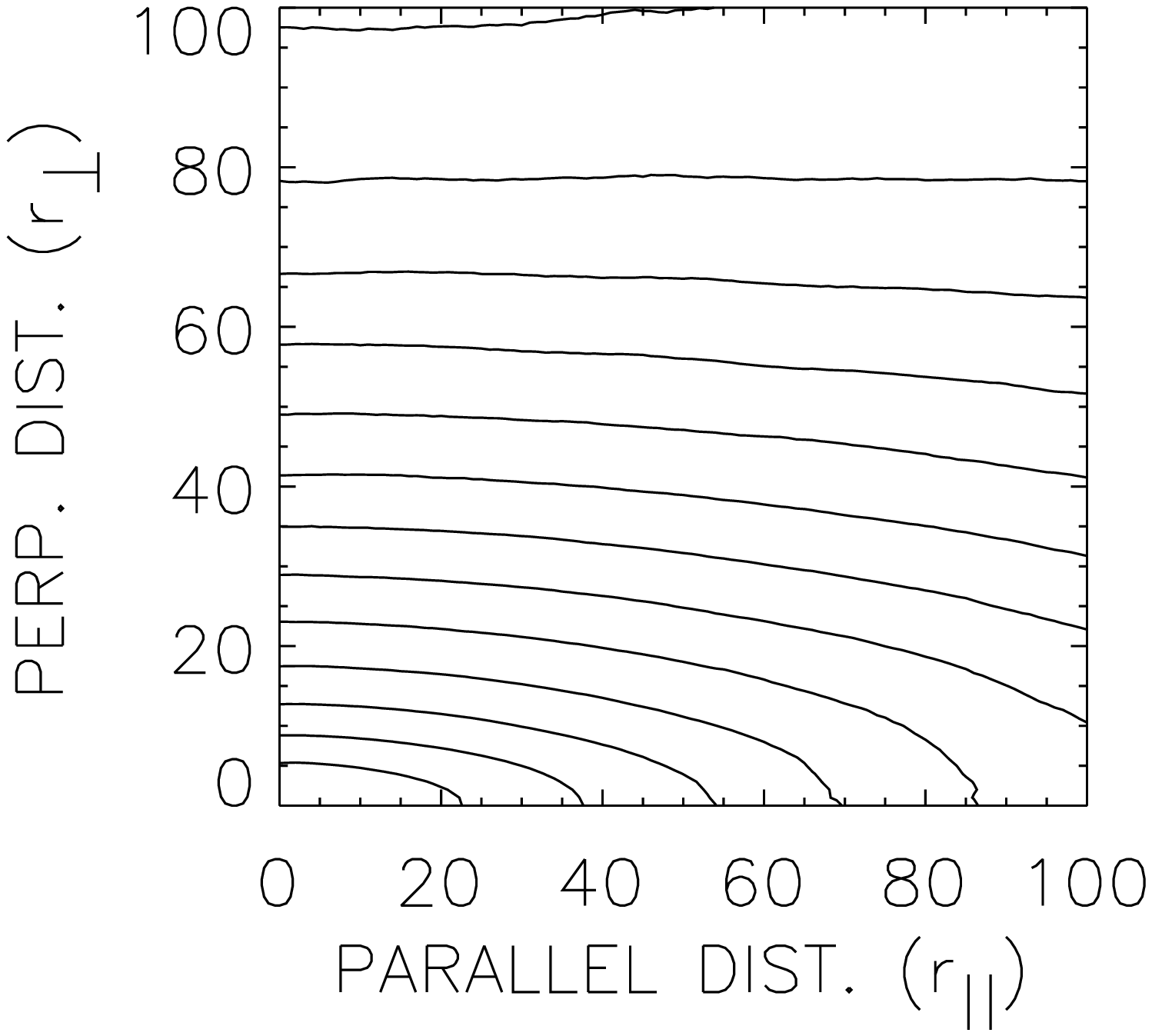}
\=
\includegraphics[width=.33\textwidth]{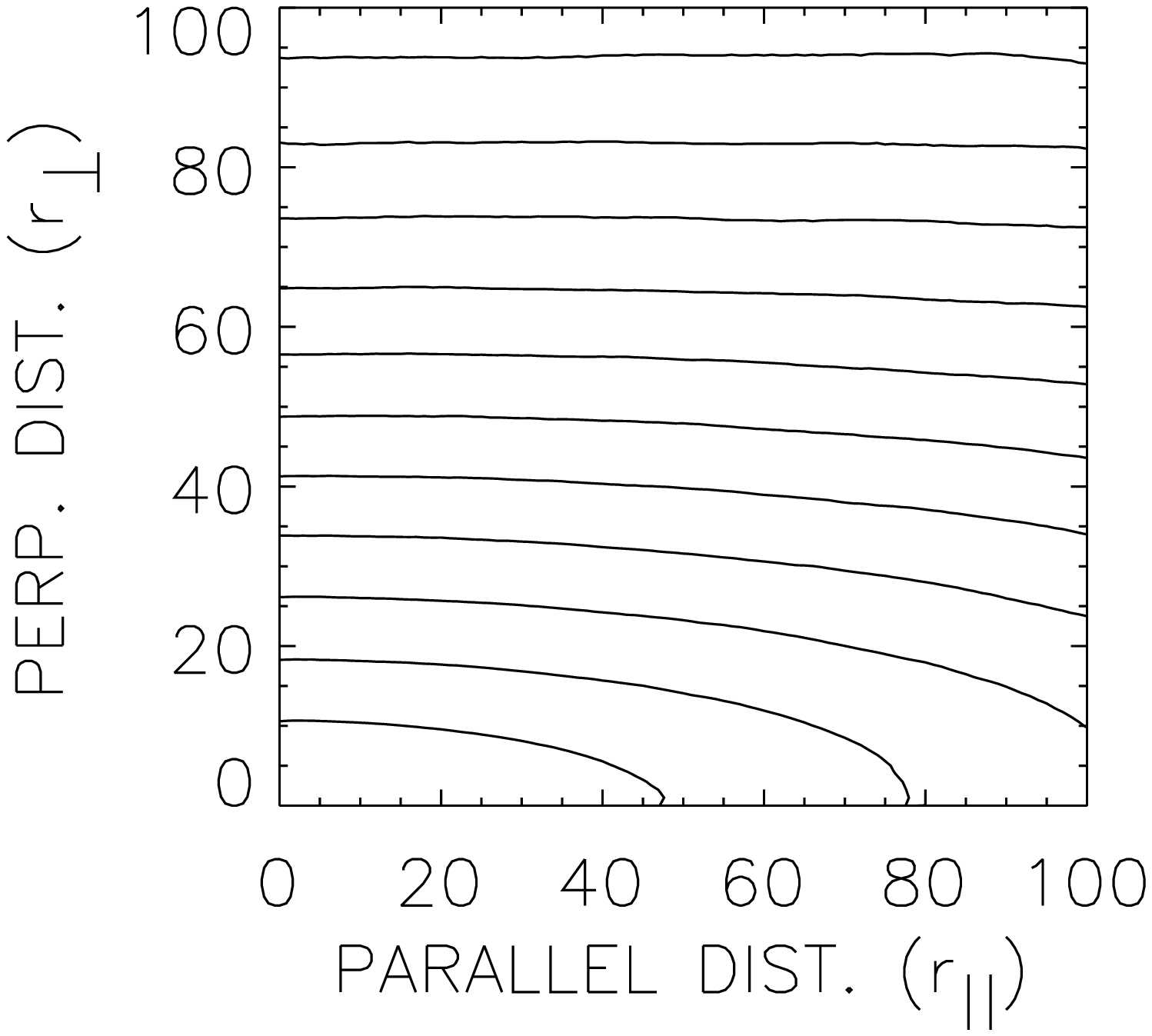}
\=
\includegraphics[width=.33\textwidth]{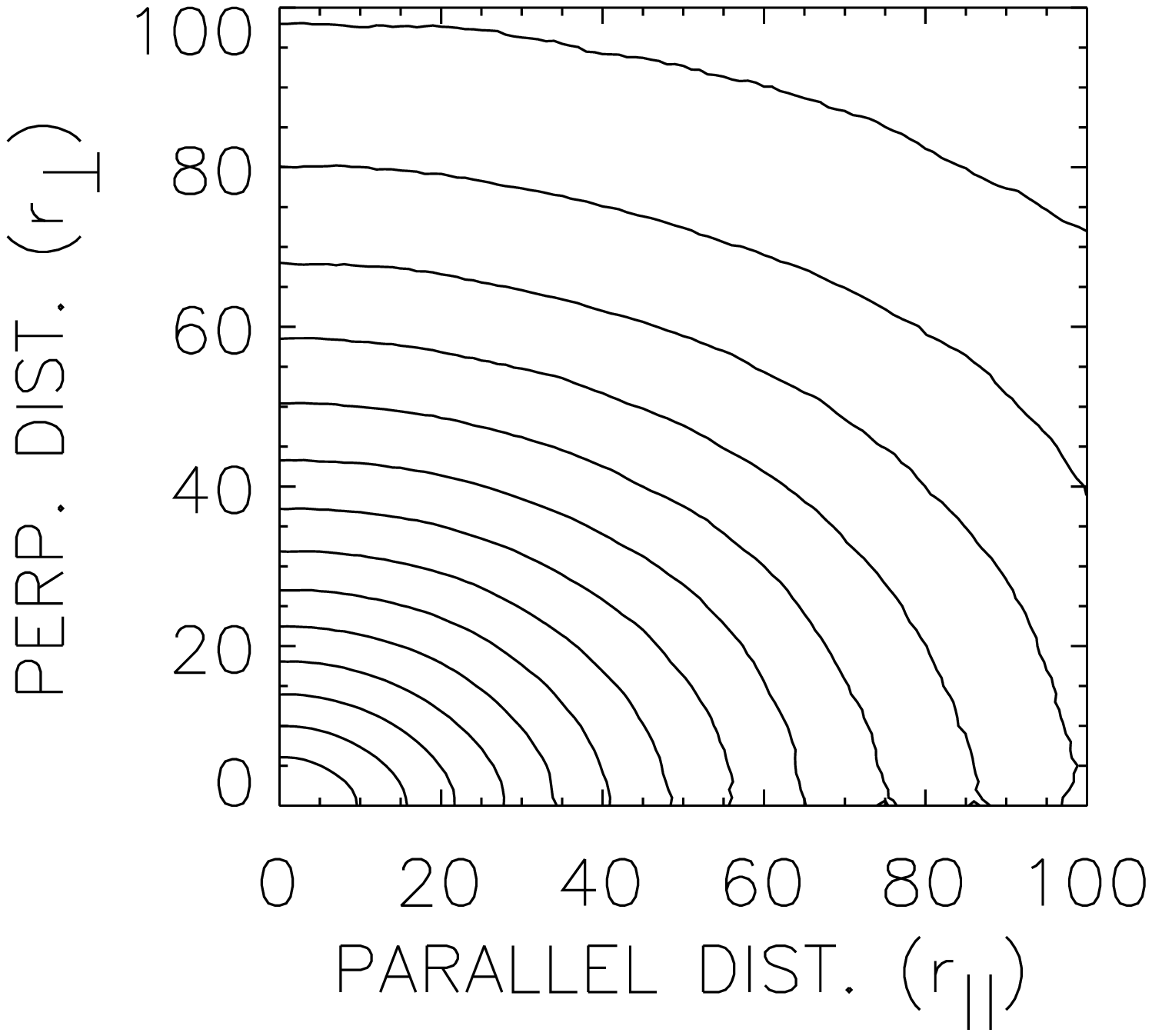}
\\   
 \hspace{23mm} (b) \=  \hspace{39mm} (d) \hspace{40mm}  \=   (f)
\end{tabbing}
  \caption{          $M_s\sim 10$, $M_A\sim 0.5$, $\beta\sim 0.01$,
           and $512^3$ grid points.
           Solid lines are for velocity spectra and
           dotted for magnetic field.
          (a) Spectra of Alfv\'en modes follow a Kolmogorov-like power
              law.
          (b) Eddy shapes
              (contours of same second-order structure function, $SF_2$)
              for velocity of Alfv\'en modes
              shows anisotropy similar to the GS95.
         ($r_{\|}\propto r_{\perp}^{2/3}$ or $k_{\|}\propto
                                              k_{\perp}^{2/3}$).
              The structure functions are measured in directions
              perpendicular or
              parallel to the local mean magnetic field in real space.
              We obtain real-space velocity and magnetic fields
              by inverse Fourier transform of
              the projected fields.
          (c) Spectra of slow modes are a bit steeper than
              the Kolmogorov-like power law. They closer to $k^{-2}$.
          (d) Slow mode velocity shows anisotropy similar to the GS95.
              We obtain contours of equal $SF_2$ directly in real space
              without going through the projection method,
              assuming slow mode velocity is nearly parallel to local
              mean magnetic field in low $\beta$ plasmas.
          (e) Spectra of fast modes show a possible departure from
              the Kolmogorov spectrum or the IK spectrum.
          (f) The magnetic $SF_2$ of
              fast modes shows isotropy. 
    } 
\label{fig_512}
\end{figure*}

\section{Density statistics}
Density at low Mach numbers follow the GS95 scaling when the driving
is incompressible (CL03). However, CL03 showed that this scaling substantially
changes for high Mach numbers. 
Our high resolution results in Fig.~3 confirm the CL03 finding 
that at high Mach numbers
density fluctuations get isotropic. Moreover, our present studies show
that the spectrum of density gets substantially
 {\it flatter}
than the GS95 one (see Fig.~4).
 Note, that a model of random shocks would produce
a spectrum {\it steeper} than the GS95 one. A possible origin of the 
flat spectrum is the superAlfvenic perturbations created by fast modes
within density perturbations originated from slow modes. This particular
regime is clearly identified in a review by CLV03 (see Fig.~9 therein). 
It may also be related to the regime
of superAlfvenic turbulence which is
discussed in Norlund \& Podoan (2003).
However, alternative explanations of the shallow density fluctuations
exist and our ongoing work should clarify which process is actually
responsible for the unusual density scaling that we observe.

The flat spectrum of density perturbations in high Mach number turbulence
is definitely a very interesting phenomenon. It does affect the properties
of interstellar medium, including the origin of the small scale clumps
that are ubiquitous in the diffuse cold gas. This flat density spectrum
should affect the star formation processes. 
A study in Boldyrev {\it et al.}~(2002) indeed testifies that the density spectrum is flat in molecular
clouds. However, a more systematic study is necessary.

\begin{figure*}
\begin{tabbing}
\includegraphics[width=.49\textwidth]{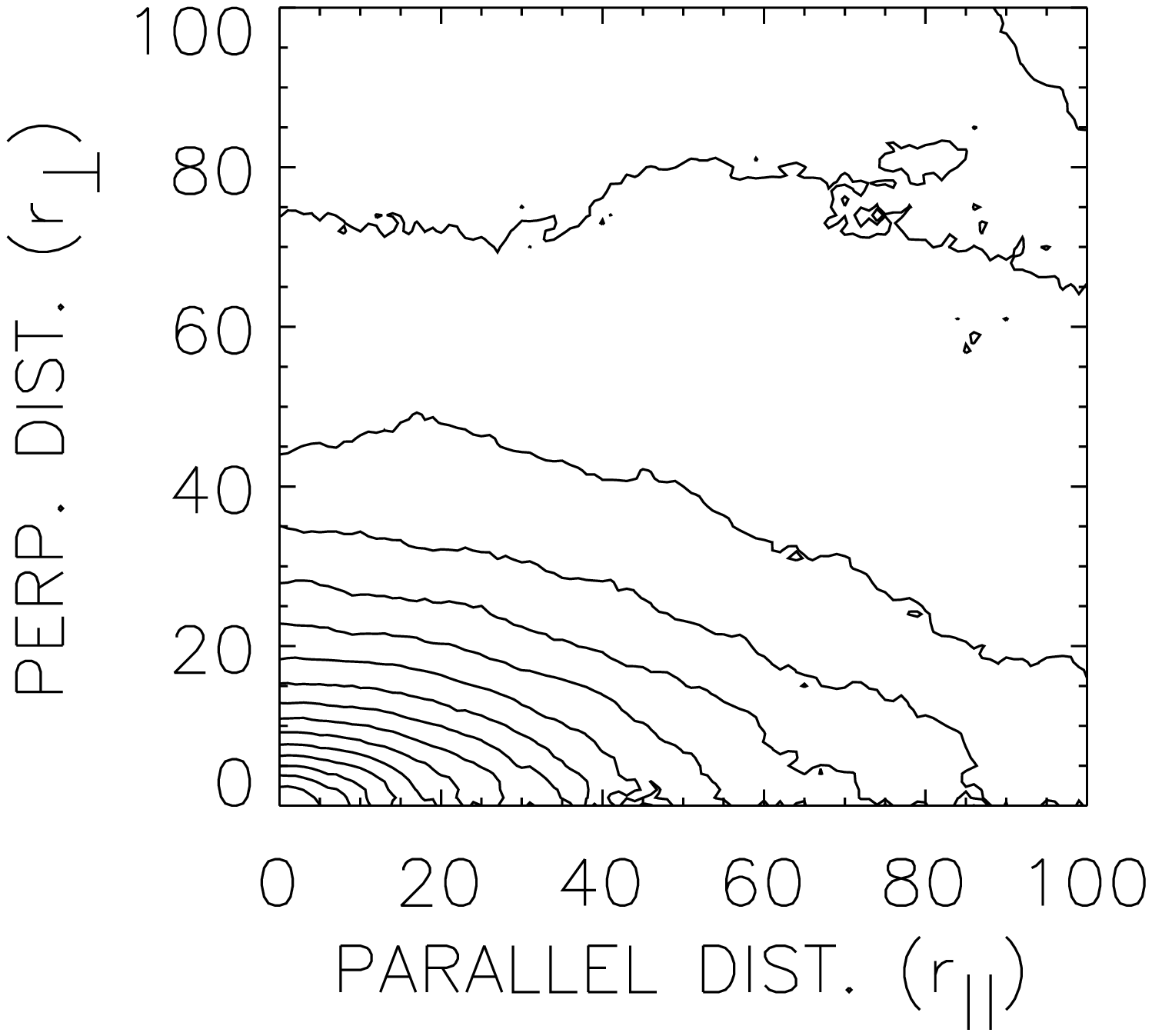}
\=  
\includegraphics[width=.49\textwidth]{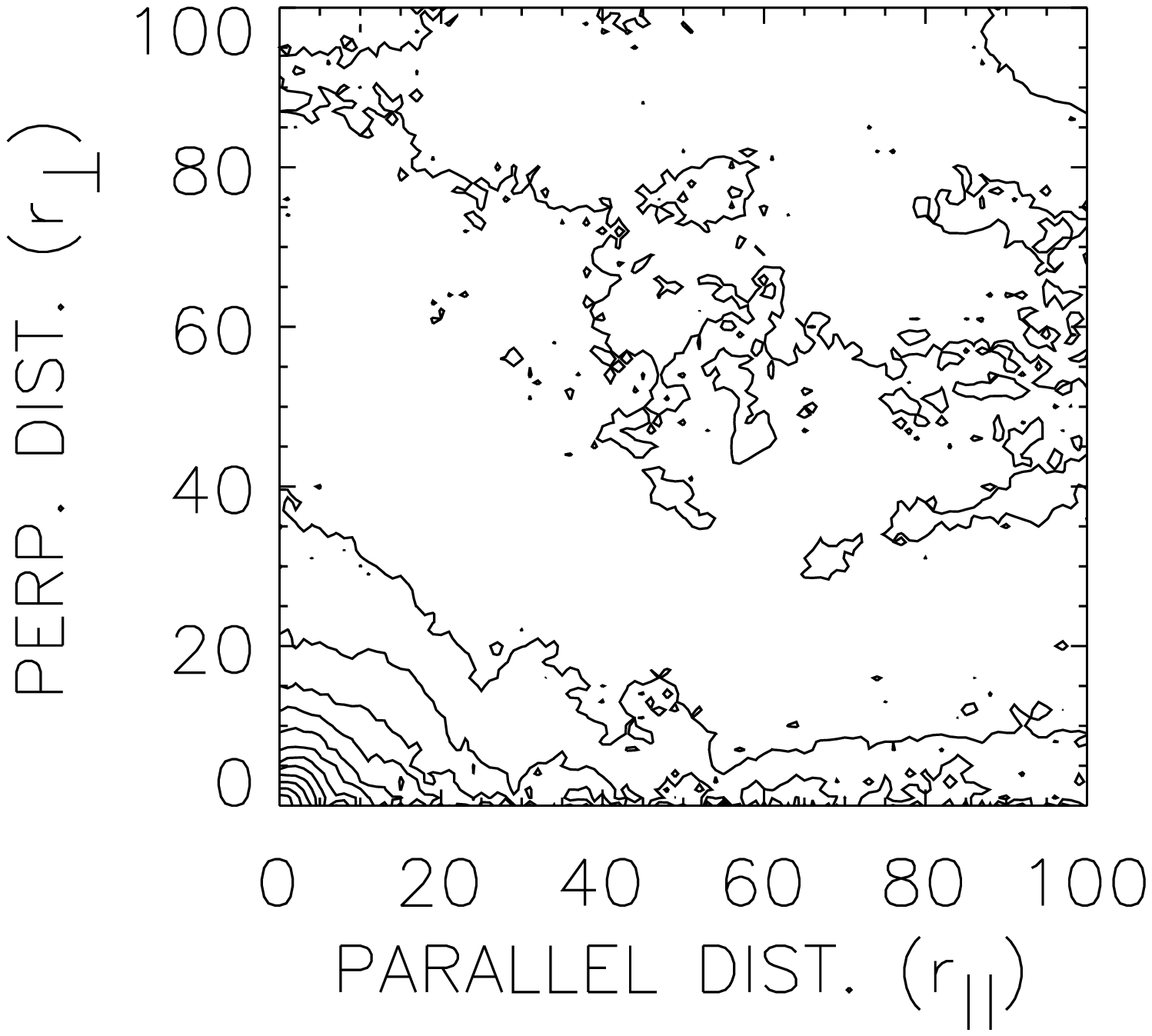}
\\   
 \hspace{38mm} (a) \= \hspace{61mm}  (b) 
\end{tabbing}
  \caption{ (a) Density structures for $M_s \sim 3$.
(b) Density structures for $M_s \sim 10$.
    Density gets isotropic as the Mach number, $M_s$, increases.
}
\label{fig_roconto}
\end{figure*}

\begin{figure}
\begin{tabbing}
\includegraphics[width=.49\textwidth]{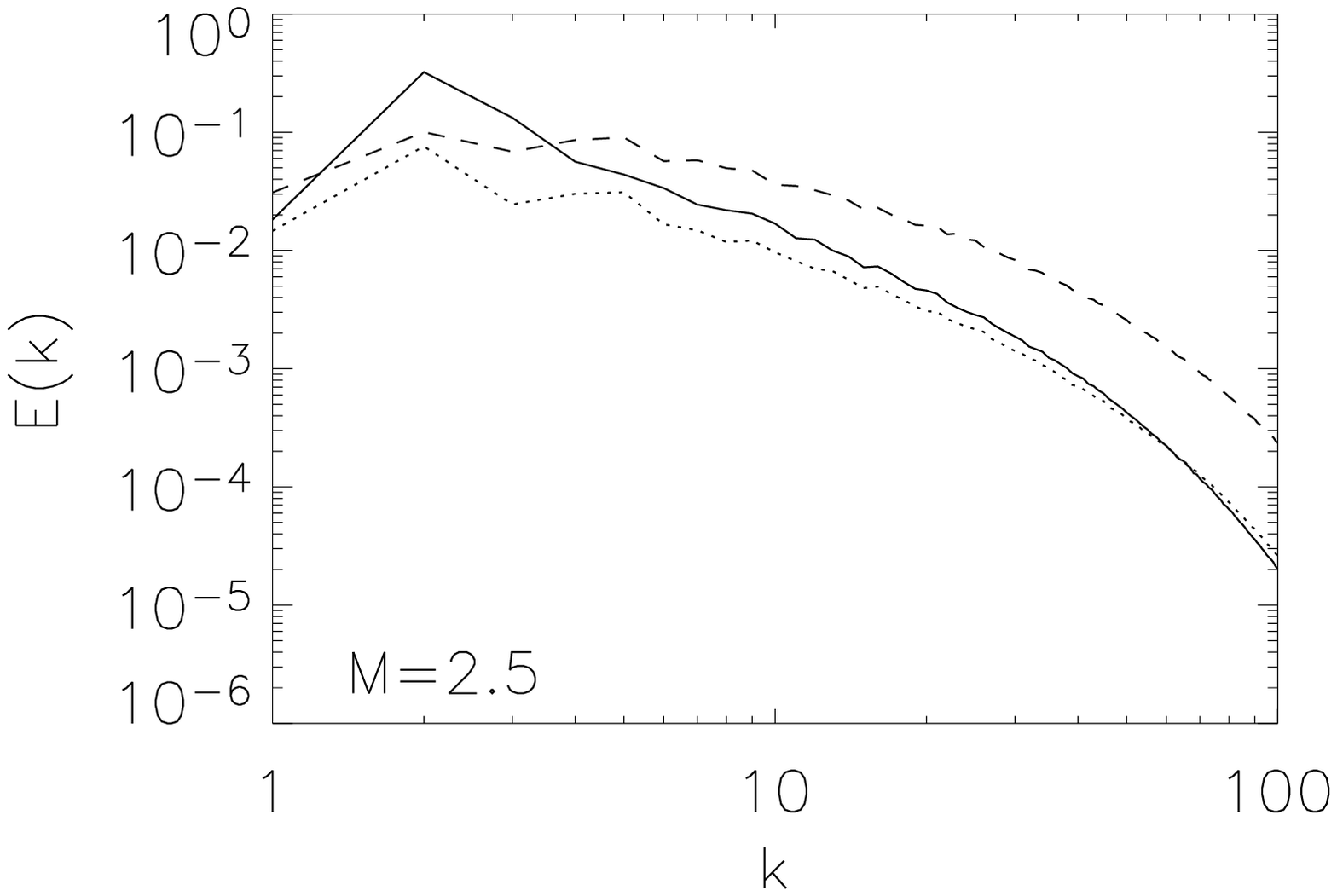}
\=  
\includegraphics[width=.49\textwidth]{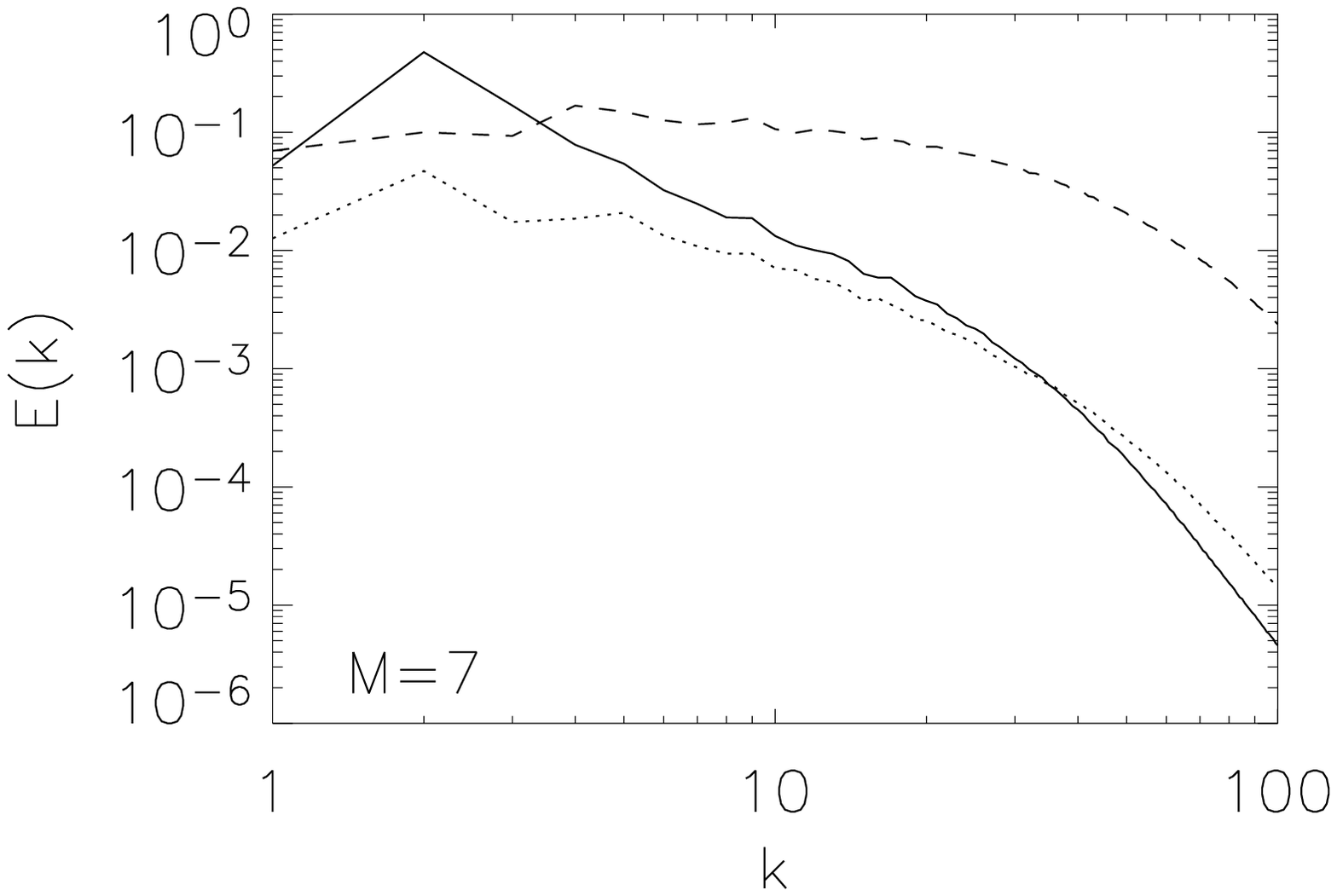}
\\   
 \hspace{36mm} (a) \= \hspace{61mm}  (b) 
\end{tabbing}
    \caption{(a) Spectra for $M_s\sim$ 3 at t=3.7.
             (b) Spectra for $M_s\sim$ 10 at t=6.0.
    Dashed lines=density; solid=velocity; dotted=magnetic field.
    Density spectrum gets flatter  
    as the Mach number, $M_s$, increases.
}
    \label{fig:s2}
\end{figure}

\section{Discussion}

In the paper above we have studied scaling relations for Alfven,
fast and slow modes. First question that comes to mind is how good
is to speak about different modes in strong MHD turbulence. It is
well known, for instance, that for strong perturbations various linear
modes are coupled. Well, our work shows that the coupling is indeed
a very important phenomenon at the large scales where the {\it total}
Mach number $M_{tot}$, which normalizes the velocity perturbations 
by the sum of
Alfvenic and sound velocities, is of the order of unity. At smaller
scales when $M_{tot}$ becomes less than unity the perturbations that
we identify through our separation of mode procedure develop without
much interaction between fast and Alfven modes. This can be understood
if we take into account that Alfven modes cascade just over one eddy
turnover time. Therefore there is not enough time for the non-linear
interaction between fast and Alfven modes to take place. At the same
time, the shearing of slow modes by Alfvenic turbulence does take place
over a turnover time. Therefore the slow modes mimic Alfven mode anisotropy.
A new effect that we start observing at high resolution is steepening
of the slow mode spectrum. This is indicative that while shearing the
Alfven modes create shocks within the gas associated with slow modes.
This is not so surprising as the Alfven velocity is smaller within
the slow modes in low beta medium.  

We have not studied here a case of turbulence that is globally superAlfvenic.
This is motivated by our belief that unless magnetic field is dynamically
important the turbulence will behave like ordinary hydro turbulence. At the
scale when the Alfven velocity gets of the order of the perturbation velocity,
we should encounter turbulence that is studied here.

Why would we care about those scalings? How wrong is it to use Kolmogorov
scalings instead? Dynamics, chemistry and physics of molecular clouds
(see Falgarone 1999) 
presents a complex of problems for which the exact scalings may be
required to a different degree. 
If we talk about dynamics of interstellar dust or
propagation of
cosmic rays, one {\it must} account for the actual scalings and couplings
of different modes (see reviews by 
Lazarian \& Yan 2003, Lazarian {\it et al.}~2003). 
There are other problems,
e.g. turbulent heat transport where the exact scaling of modes seems
to be less important (Cho {\it et al.}~2003). 

\section{Conclusion}
In the brief, we have 
studied statistics of compressible MHD turbulence
with high resolution numerical simulation for
highly ($M_s\sim 10$) supersonic low-$\beta$ case.
We provided the decomposition
of turbulence into Alfven, slow and fast modes.
We have found that GS95
scaling is valid for {\it Alfven modes}:
$$
   \mbox{ Alfv\'{e}n:~}  E^A(k)  \propto k^{-5/3}, 
                        ~~~k_{\|} \propto k_{\perp}^{2/3}. 
$$
{\it Slow modes} also follows the GS95 anisotropy for
highly supersonic low-$\beta$ case:
$$
   \mbox{ Slow:~~~}    
                       k_{\|} \propto k_{\perp}^{2/3}.  
$$
The velocity spectrum of 
slow modes tends to be steeper, which may be related
to the formation of shocks.

Density shows anisotropy for the mildly supersonic case.
But, for the highly supersonic case, density shows more or less isotropic
structures.
We suspect that
shock formation is responsible for the isotropization of density.
Our present studies show
that the spectrum of density gets substantially
 {\it flatter}
than the GS95 one (see Fig.~4).

\section{Acknowledgments}
The authors are grateful to CTR for hospitality. AL acknowledge
a support by the NSF Center for Magnetic Self-Organization in
the Laboratory and Astrophysical Plasmas.


\begin{thebibliography}{}

\bibitem[\protect\citeauthoryear%
{Armstrong, Rickett, \& Spangler}{Armstrong et al.}{1995}]{ArmRS95}
 Armstrong,ÊJ.ÊW., Rickett,ÊB.ÊJ., \& Spangler,ÊS.ÊR., 1995,
Electron density power spectrum in the local interstellar medium,
 Astrophysical J., 443, 209-221


\bibitem[\protect\citeauthoryear%
{Boldyrev, Nordlund, \& Padoan}{Boldyrev et al.}{2002}]{BolNP02b}
Boldyrev, S., Nordlund, \AA., \& Padoan P. 2002, 
Supersonic Turbulence and Structure of Interstellar Molecular Clouds,
Phys. Rev. Lett. 89, 031102

\bibitem[\protect\citeauthoryear%
{Cho \& Lazarian}{2002a}]{ChoL02a}
 Cho, J., Lazarian, A. 2002, 
   Compressible Sub-AlfvŽnic MHD Turbulence in Low-$\beta$ Plasmas,
   Phy. Rev. Lett., 88, 245001 (CL02)


\bibitem[\protect\citeauthoryear%
{Cho \& Lazarian}{2003a}]{ChoL03a} 
      Cho, J., Lazarian, A. 2003, 
      Compressible magnetohydrodynamic turbulence: mode coupling, scaling relations, anisotropy, viscosity-damped regime and astrophysical implications,
      MNRAS, 345, 325-339 (CL03)

\bibitem[\protect\citeauthoryear%
{Cho \& Lazarian}{2004}]{ChoL04} 
      Cho, J., Lazarian, A. 2004, 
     Generation of Compressible Modes in MHD Turbulence,
     Theo. Comp. Fluid Dyn., accepted 
    (http://xxx.lanl.gov/abs/astro-ph/0301462) 


\bibitem[\protect\citeauthoryear%
{Cho et al.}{2003}]{ChoLH03}
Cho, J., Lazarian, A., Honein, A., Knaepen, B., Kassinos, S., \& Moin, P.
2003, 
Thermal Conduction in Magnetized Turbulent Gas,
ApJ, 589, L77-L80

\bibitem[\protect\citeauthoryear%
{Cho, Lazarian \& Vishniac}{Cho et al.}{2002a}]{ChoLV02b}
 Cho, J., Lazarian, A., \& Vishniac, E. 2002a, 
Simulations of Magnetohydrodynamic Turbulence in a Strongly Magnetized Medium,
ApJ, 564, 291-301  (CLV02a)


\bibitem[\protect\citeauthoryear%
{Cho, Lazarian \& Vishniac}{Cho et al.}{2003a}]{ChoLV03a}
 Cho, J., Lazarian, A., \& Vishniac, E. 2003, in
      {\it Turbulence and Magnetic Fields in Astrophysics},
      eds. E. Falgarone \& T. Passot (Springer LNP), p56-98
      (astro-ph/0205286)  (CLV03)


\bibitem[\protect\citeauthoryear%
{Cho \& Vishniac}{2000b}]{ChoV00b}
 Cho, J. \& Vishniac, E. 2000, 
 The Anisotropy of Magnetohydrodynamic AlfvŽnic Turbulence,
 ApJ, 539, 273-282



\bibitem[\protect\citeauthoryear%
{Falgarone}{1999}]{Fal99} Falgarone, E. 1999,
The Intermittent Dissipation of Turbulence: is it Observed in the Interstellar Medium?,
    in {\it Interstellar Turbulence}, ed. by J. Franco,
                A. Carraminana, CUP,  p.132

\bibitem[\protect\citeauthoryear%
{Goldreich \& Sridhar}{1995}]{GolS95}
 Goldreich, P. \& Sridhar, S. 1995, 
 Toward a theory of interstellar turbulence. 2: Strong alfvenic turbulence,
 ApJ, 438, 763-775 (GS95)


\bibitem[\protect\citeauthoryear%
{Higdon}{1984}]{Hig84}
 Higdon, J. C. 1984, 
 Density fluctuations in the interstellar medium: Evidence for anisotropic magnetogasdynamic turbulence. I - Model and astrophysical sites,
 ApJ, 285, 109-123



\bibitem[\protect\citeauthoryear%
{Lazarian et al.}{2003}]{LazPY03}
Lazarian, A.,  Petrosian, V., Yan, H., \& Cho, J. 2003, preprint 
(astro-ph/0301181)

\bibitem[\protect\citeauthoryear%
{Lazarian \& Prunet}{2002}]{LazP02}
 Lazarian, A. \& Prunet, S. 2002,
 Polarized microwave emission from dust, 
 in {\it Astrophysical Polarized Backgrounds}, 
                ed. by S. Cecchini, S. Cortiglioni, R. Sault, C. Sbarra,
         (Melville, NY; AIP, Vol. 109) pp.32-43 (astro-ph/0111214)


\bibitem[\protect\citeauthoryear%
{Lazarian \& Yan}{2002}]{LazY02}
 Lazarian, A. \& Yan, H. 2003, 
 Translational Velocities and Rotational Rates of Interstellar Dust  
 Grains,
 in ``Astrophysical Dust''
eds. A. Witt \& B. Draine, APS, in press (astro-ph/0311370)


\bibitem[\protect\citeauthoryear%
{Lithwick \& Goldreich}{2001}]{LitG01}
 Lithwick, Y. \& Goldreich, P. 2001, 
 Compressible Magnetohydrodynamic Turbulence in Interstellar Plasmas,
 ApJ, 562, 279-296


\bibitem[\protect\citeauthoryear%
{Maron \& Goldreich}{2001}]{MarG01}
 Maron, J. \& Goldreich, P. 2001, 
 Simulations of Incompressible Magnetohydrodynamic Turbulence, 
 ApJ, 554, 1175-1196


\bibitem[\protect\citeauthoryear%
{Matthaeus \& Brown}{1988}]{MatB88}
Matthaeus, W.~H. \& Brown, M.~R. 1988, 
Nearly incompressible magnetohydrodynamics at low Mach number,
Phys. Fluids, 31(12), 3634-3644

\bibitem[\protect\citeauthoryear%
{Matthaeus et al.}{1996}]{MatGO96}
Matthaeus, W.~H., Ghosh, S., Oughton, S., \& Roberts, D.~A. 1996, 
Anisotropic three-dimensional MHD turbulence,
J.~Geophysical Res., 101(A4), 7619-7630




\bibitem[\protect\citeauthoryear%
{Monin \& Yaglom}{1975}]{} 
    Monin, A.S., \& Yaglom, A.M. 1975, 
 Statistical Fluid Mechanics: Mechanics of Turbulence, vol. 2, The MIT Press



\bibitem[\protect\citeauthoryear%
{Nordlund \& padoan}{2003}]{}
      Norlund, A., Podoan, P. 2003, 
      Star Formation and the Initial Mass Function,
      in
      {\it Turbulence and Magnetic Fields in Astrophysics},
      eds. E. Falgarone \& T. Passot (Springer LNP), p.271-298


\bibitem[\protect\citeauthoryear%
{Pouquet}{1999}]{}  Pouquet, A. 1999,
    An Introduction to Compressible MHD Turbulence,
    in {\it Interstellar Turbulence}, p.87

\bibitem[\protect\citeauthoryear%
{Shebalin, Matthaeus, \& Montgomery}{Shebalin et al.}{1983}]{SheMM83}
Shebalin, J.~V., Matthaeus, W.~H., \& Montgomery, D.~C. 1983, 
    Anisotropy in MHD turbulence due to a mean magnetic field,
                 J.~Plasma~Phys.,  29, 525-547








\bibitem[\protect\citeauthoryear%
{Zank \& Matthaeus}{1993}]{ZanM93}
Zank, G. P. \& Matthaeus, W. H. 1993, 
Nearly incompressible fluids. II - Magnetohydrodynamics, turbulence, and waves,
Phys.~Fluids A, 5(1), 257-273


  


\end{thebibliography}
\end{document}